\documentclass[prd,showpacs, twocolumn,amsmath,amssymb]{revtex4}

\usepackage{amssymb}
\usepackage{graphicx}
\usepackage{dcolumn}
\usepackage{bm}
\usepackage{epsfig}
\usepackage{amsfonts}
\usepackage{keyval,graphicx}
\usepackage{textcomp,wasysym}
\usepackage{subfigure}
\usepackage[none]{hyphenat}
\usepackage{xcolor}

\begin{document}

\newcommand{\chisq}[1]{$\chi^{2}_{#1}$}
\newcommand{\etap}{\eta^{\prime}}
\newcommand{\pip}{\pi^{+}}
\newcommand{\pim}{\pi^{-}}
\newcommand{\piz}{\pi^{0}}
\newcommand{\rhoz}{\rho^{0}}
\newcommand{\az}{a_{0}(980)}
\newcommand{\fz}{f_{0}(980)}
\newcommand{\pipm}{\pi^{\pm}}
\newcommand{\psip}{\psi^{\prime}}
\newcommand{\jpsi}{J/\psi}
\newcommand{\ar}{\rightarrow}
\newcommand{\GeV}{GeV/$c^2$}
\newcommand{\MeV}{MeV/$c^2$}
\newcommand{\br}[1]{\mathcal{B}(#1)}
\newcommand{\cinst}[2]{$^{\mathrm{#1}}$~#2\par}
\newcommand{\crefi}[1]{$^{\mathrm{#1}}$}
\newcommand{\crefii}[2]{$^{\mathrm{#1,#2}}$}
\newcommand{\crefiii}[3]{$^{\mathrm{#1,#2,#3}}$}
\newcommand{\HRule}{\rule{0.5\linewidth}{0.5mm}}


\title{\Large \boldmath \bf Search for $\eta$ and $\eta^\prime\to \pi^+ e^- \bar{\nu}_e +c.c.$
decays in $\jpsi \to \phi \eta$ and $\phi \eta^\prime$ }

\author{\small
M.~Ablikim$^{1}$, M.~N.~Achasov$^{6}$, O.~Albayrak$^{3}$,
D.~J.~Ambrose$^{39}$, F.~F.~An$^{1}$, Q.~An$^{40}$, J.~Z.~Bai$^{1}$,
R.~Baldini Ferroli$^{17A}$, Y.~Ban$^{26}$, J.~Becker$^{2}$
% ...
\\
\vspace{0.2cm}
(BESIII Collaboration)\\
\vspace{0.2cm} {\it
$^{1}$ Institute of High Energy Physics, Beijing 100049, People's Republic of China\\
$^{2}$ Bochum Ruhr-University, D-44780 Bochum, Germany\\
$^{3}$ Carnegie Mellon University, Pittsburgh, Pennsylvania 15213, USA\\
% ...
$^{6}$ G.I. Budker Institute of Nuclear Physics SB RAS (BINP), Novosibirsk 630090, Russia\\
% ...
$^{17}$ (A)INFN Laboratori Nazionali di Frascati, I-00044, Frascati, Italy; (B)INFN and University of Perugia, I-06100, Perugia, Italy\\
% ...
$^{26}$ Peking University, Beijing 100871, People's Republic of China\\
% ...
$^{39}$ University of Rochester, Rochester, New York 14627, USA\\
$^{40}$ University of Science and Technology of China, Hefei 230026, People's Republic of China\\
% ...
}}

\vspace{0.4cm}


\collaboration{${\it BESIII}$  Collaboration}
\date{\today}


\date{\today}

\begin{abstract}
Using a sample of 225.3 million $\jpsi$ events collected with the
BESIII detector at the BEPCII $e^+e^-$ collider in 2009, searches
for the decays of $\eta$ and $\eta^\prime\to\pi^+ e^- \bar{\nu}_e
+c.c.$ in $\jpsi \to \phi \eta$ and $\phi\eta^\prime$ are performed.
The $\phi$ signals, which are reconstructed in $K^+K^-$ final
states, are used to tag  $\eta$ and $\eta^\prime$ semileptonic
decays. No signals are observed for either $\eta$ or $\eta^\prime$,
and upper limits at the 90\% confidence level are determined to be
$7.3\times 10^{-4}$ and $5.0\times 10^{-4}$ for the ratios
$\frac{{\mathcal B}(\eta\to \pi^+ e^- \bar{\nu}_e +c.c.)}{{\mathcal
B}(\eta \to \pip\pim\piz)}$ and $\frac{{\mathcal B}(\eta^\prime\to
\pi^+ e^-\bar{\nu}_e +c.c.)}{{\mathcal B}(\eta^\prime \to
\pip\pim\eta)}$, respectively. These are the first upper limit
values determined for $\eta$ and $\eta^\prime$ semileptonic weak
decays.
\end{abstract}


\maketitle


%
\section{Introduction}
\label{sec:intro}

 Weak decays of quarkonium states such as $\eta$, $\eta^\prime$, $J/\psi$ and $\Upsilon$,
 etc., offer a window into what may lie beyond the standard model
 (SM)~\cite{fayet-1,fayet-2,fayet-3,besii,cleo-1,cleo-2}. The reason for the expected sensitivity is that the rates of the quarkonium weak decays are expected to be tiny in
 the framework of the SM~\cite{lihb}. As originally pointed out by Singer~\cite{singer},
 the weak decays $\eta \to \pi^+ l^- \bar{\nu}_l $ ($l=e$, $\mu$, and charge conjugate
state implicitly included) are purely second class
with a vector-type coupling in the SM (see Ref.~\cite{weinberg} for
the definition of the second class current), and hence vanish in the
limit of exact isospin symmetry. They occur in the SM in first order
in the weak interaction, but only due to $G$-parity breaking
effects, i.e. due to electromagnetic corrections and the
mass-difference of the $u$- and
$d$-quarks~\cite{kun,peter,neufeld,bramon,ng,lucha,kim,shabalin}.
 For $\eta$ semileptonic weak decays, a one-loop calculation was performed
 in chiral perturbation theory within the SM, including a systematic treatment of the electromagnetic
contributions to $O(e^2 p^2)$ ($e$ and $p$ are electromagnetic
coupling and typical momentum transfer in the decay as defined in
Ref.~\cite{neufeld}), and a rather accurate upper
 bound for the branching fraction of $\eta \to \pi^+ l^-
 \bar{\nu}_l$  is predicted to be $2.6 \times 10^{-13}$. Therefore, any observation
of $\eta\to \pi l \nu_l$  violating this bound would be a clear
indication for new physics beyond the SM.

The decays $\eta\to \pi l \nu_l$ can be used to probe some types of
possible new charged current interactions~\cite{peter,kun}. A rather
old suggestion would be the introduction of a new second class
vector current for the $\eta \to \pi$ transition~\cite{lucha}.
Scalar-type charged current four-fermion interactions can arise in
gauge theories for example from the exchange of charged Higgs bosons
in the two-Higgs-doublet model~\cite{lee,2hdm}. Also light
leptoquarks~\cite{barr}, occurring naturally in grand unified
theories and composite models, may enhance the  $\eta\to \pi l
\nu_l$ branching faction considerably~\cite{wyler}.  For example, by
considering scalar or vector type interaction, the branching
fraction of $\eta \to \pi^+ l^-
 \bar{\nu}_l$ was estimated to be $ 10^{-8}-10^{-9}$~\cite{peter-2,peter-3}, which is a few order of magnitudes higher than that in the SM.
 Therefore, searches for the $\eta \to \pi^+ l^-
 \bar{\nu}_l$ and  $\eta^\prime \to \pi^+ l^-
 \bar{\nu}_l$  at the branching fractions level of $ 10^{-8}-10^{-9}$ and below will provide information on the new physics beyond the SM.
   At present
there is no experimental information on the decays $\eta\to \pi l
\nu_l$. In this paper, we present measurements of branching
fractions of $\eta$ and $\eta^\prime\to\pi^+ e^- \bar{\nu}_e $
decays. This analysis is based on
 $(225.3\pm 2.8)\times 10^6$
$J/\psi$ events~\cite{njpsi}, accumulated with the Beijing
Spectrometer III (BESIII) detector~\cite{besnim}, at the Beijing
Electron Positron Collider II (BEPCII).

\section{The BESIII Experiment and MC simulation}
\label{sec:detector}

BEPCII/BESIII~\cite{besnim} is a major upgrade of the BESII
experiment at the BEPC accelerator. The design peak luminosity of
the double-ring $e^+e^-$ collider, \nohyphens{BEPCII}, is $10^{33}$
cm$^{-2}$s$^{-1}$ at the center-of-mass energy of 3770 MeV.
  The BESIII detector has a geometrical
acceptance of 93\% of $4\pi$ and consists of four main components:
(1) a small-celled, helium-based main drift chamber (MDC) with 43
layers, which provides measurements of ionization energy loss
($dE/dx$). The average single wire resolution is 135 $\mu$m, and the
momentum resolution for charged particles with momenta of 1 GeV/$c$
in a 1 T magnetic field is 0.5\%; (2) an electromagnetic calorimeter
(EMC) made of 6240 CsI (Tl) crystals arranged in a cylindrical shape
(barrel) plus two end-caps. For 1.0 GeV photons, the energy
resolution is 2.5\% in the barrel and 5\% in the end-caps, and the
position resolution is 6 mm in the barrel and 9 mm in the end-caps;
(3) a time-of-flight system (TOF) for particle identification (PID)
composed of a barrel part made of two layers with 88 pieces of 5 cm
thick and 2.4 m long plastic scintillators in each layer, and two
end-caps with 96 fan-shaped, 5 cm thick, plastic scintillators in
each end-cap. The time resolution is 80 ps in the barrel, and 110 ps
in the end-caps, corresponding to a 2$\sigma$ K/$\pi$ separation for
momenta up to about 1.0 GeV/$c$; (4) a muon chamber system made of
1000 m$^2$ of resistive-plate-chambers arranged in 9 layers in the
barrel and 8 layers in the end-caps and incorporated in the return
iron of the super-conducting magnet. The position resolution is
about 2 cm.

The optimization of the event selection and the estimation of
physics backgrounds are performed using Monte Carlo (MC) simulated
data samples. The {\sc geant4}-based simulation software {\sc
BOOST}~\cite{geant4} includes the geometric and material description
of the BESIII detectors, the detector response and digitization
models, as well as the track records of the detector running
conditions and performance. The production of the $J/\psi$ resonance
is simulated by the MC event generator {\sc kkmc}~\cite{kkmc}; the
known decay modes are generated by {\sc evtgen}~\cite{evtgen} with
branching ratios taken from the Particle Data Group (PDG)
tables~\cite{pdg}
and determined by the Lundcharm model
{\sc lundcharm}~\cite{lund} for the remaining unknown decays. The
analysis is performed in the framework of the BESIII offline
software system~\cite{soft} which takes care of the detector
calibration, event reconstruction and data storage.

\section{Data Analysis }
\label{sec:selection}

\subsection{Analyses for $\eta$ and $\eta^\prime \to \pi^+ e^- \bar{\nu}_e $  }
\label{sec:invisible:selection}

In order to detect $\eta$ and $\eta^\prime\to\pi^+ e^- \bar{\nu}_e $
decays, we use $\jpsi \to \phi \eta$ and $\phi\eta^\prime$ decays.
These two-body decays provide a very simple event topology, in which
the $\phi$ signals can be reconstructed easily and cleanly decaying
into $K^+K^-$. The reconstructed $\phi$ particles can be used to tag
$\eta$ and $\eta^\prime$ in order to allow a search for their
semileptonic decays. In addition, the $\eta$ and $\eta^\prime$
decays are easy to define in the lab system due to the strong boost
of the $\phi$ from $J/\psi$ decay.

 Charged tracks in the BESIII detector are reconstructed using
track-induced signals in the MDC. We select tracks  within $\pm10$
cm of the interaction point in the beam direction and within 1 cm in
the plane perpendicular to the beam direction. The tracks must be
within the MDC fiducial volume, $|\cos\theta| < 0.93$ ($\theta$ is
the polar angle with respect to the $e^+$ beam direction). Candidate
events require four charged tracks with net charge zero.
The TOF and $dE/dx$ information are combined to form PID confidence
levels for the $\pi$, $K$, and $e$ hypotheses; each track is
assigned to the particle type that corresponds to the hypothesis
with the highest confidence level. To suppress background from
$J/\psi \to \phi \eta~(\eta^\prime)$, where $\eta$~($\eta^\prime$)
decays into nonleptonic modes, the electron candidate is further
identified with the ratio of deposited energy in the EMC to track
momentum, $E/p$, which must be larger than 0.8. We further require
that $E/p$ should be less than 0.8 for the pion candidate to
suppress background from $J/\psi \to \phi \eta$ ($\eta \to \gamma
e^+e^-$) decay.

Showers identified as photon candidates must satisfy
fiducial and shower-quality requirements. The minimum energy is 25
MeV for EMC barrel showers ($|\cos\theta|<0.8$) and 50 MeV for
end-cap showers ($0.86<|\cos\theta|<0.92$).  To eliminate showers
produced by charged particles, a photon must be isolated from any
charged track by more than $20^\circ$ if not specified otherwise.

 Since the mass of neutrino is almost zero and it is invisible in the detectors,
 a one-constraint (1C) kinematic fit is performed to constrain the missing
mass of the reconstructed tracks to be zero, and \chisq{1C} $<200$
is required.  The 1C fit improves the resolution of recoil mass of
the $K^+K^-$ system by a factor of 2.5 for $\eta$ case or a factor
of 1.6 for $\eta^\prime$ case. After the 1C fit, the missing
momentum $P_{\text{miss}} = |\vec{P}_{\text{miss}}|$ can be
calculated; here, $\vec{P}_{\text{miss}}= -(\vec{P}_{\phi}+
\vec{P}_{\pi^+} + \vec{P}_{e^-})$ in the rest frame of $J/\psi$, and
we require that the missing momentum should be larger than $0.03$
GeV/$c$ to suppress backgrounds from final states with only four
tracks, such as $J/\psi \to \phi \pi^+\pi^-$ ($\phi\to K^+K^-$). We
count the number $N_{\text{shower}}$ of EMC showers that could
originate from a $K_L$ or a photon, and require that
$N_{\text{shower}}$ be zero in the region inside a cone of 0.3 (1.5)
rad around the direction of the missing momentum for $\jpsi \to \phi
\eta (\eta^\prime)$ [$\eta (\eta^\prime) \to \pi^+ e^- \bar{\nu}_e
$]. These requirements reject most $\eta$ and $\eta^\prime$ decays
into nonleptonic  final states. They also eliminate most backgrounds
from multi-body decays of $J/\psi \to \phi+\text{anything}$. The
different requirements on the cone angle for the $\eta$ and
$\eta^\prime$ cases are made because of the following two reasons:
firstly, in the $J/\psi \rightarrow \phi \eta (\eta^\prime)$ decays,
the booster for $\eta$ is stranger than that for $\eta^\prime$ in
the central of mass energy of $J/\psi$, which leads to a larger open
angle for the $\eta^\prime$ decay products than that for the $\eta$
decay products in the detector. Secondly, the most dangerous
backgrounds are from $\eta (\eta^\prime) \rightarrow \pi^+\pi^-
\gamma$ decay, in which one of the charged pions is mis-identified
as an electron. Meanwhile, the decay rate for $\eta^\prime
\rightarrow \pi^+\pi^- \gamma$ is more than 6 times larger than the
rate for $\eta \rightarrow \pi^+\pi^- \gamma$~\cite{pdg}.

\begin{figure}[hbtp]
  {\label{fig:subfig:a}\includegraphics[width=0.48\columnwidth]{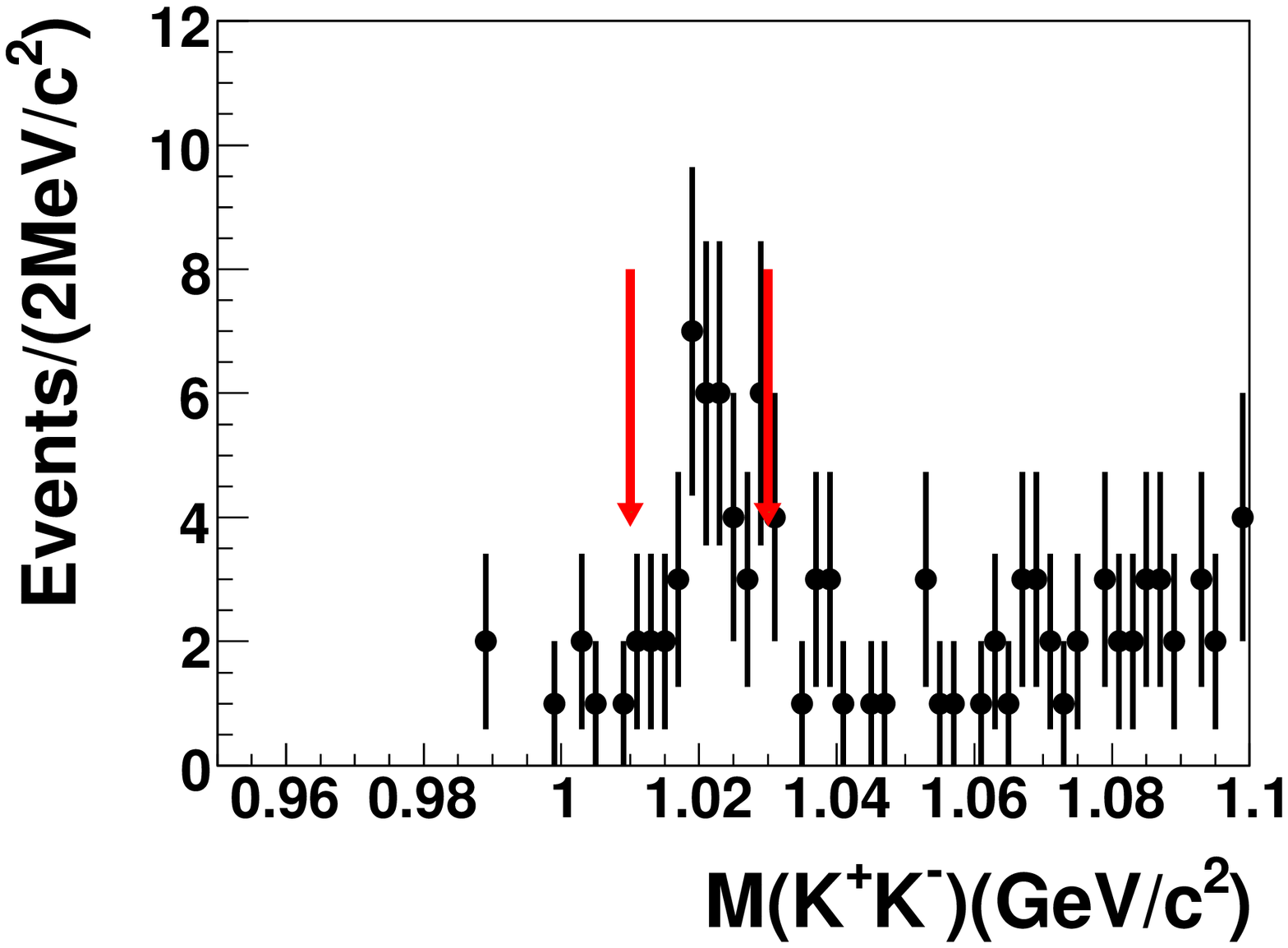}}
  \put(-90,72) {\bf \scriptsize (a)}
  {\label{fig:subfig:b}\includegraphics[width=0.48\columnwidth]{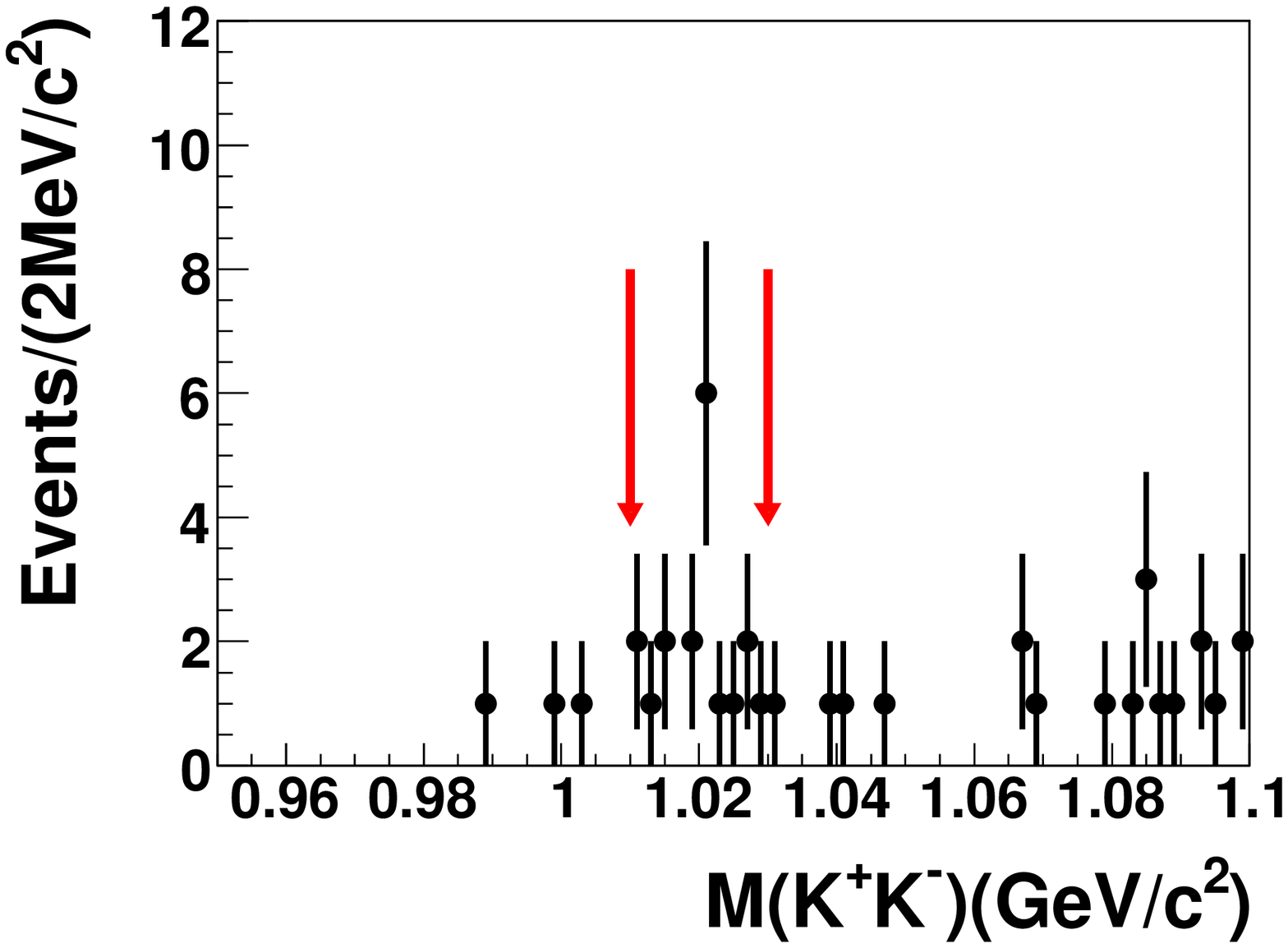}}
  \put(-90,72) {\bf \scriptsize (b)}
\caption{The $m_{K^+K^-}$ distributions of candidate events:
 (a) for $J/\psi \to \phi \eta$; (b) for $J/\psi \to \phi
 \eta^\prime$. The arrows on the plots indicate the signal region of $\phi$
candidates.} \label{phie}
\end{figure}

\begin{figure}[hbtp]
  {\label{fig:subfig:a}\includegraphics[width=0.48\columnwidth]{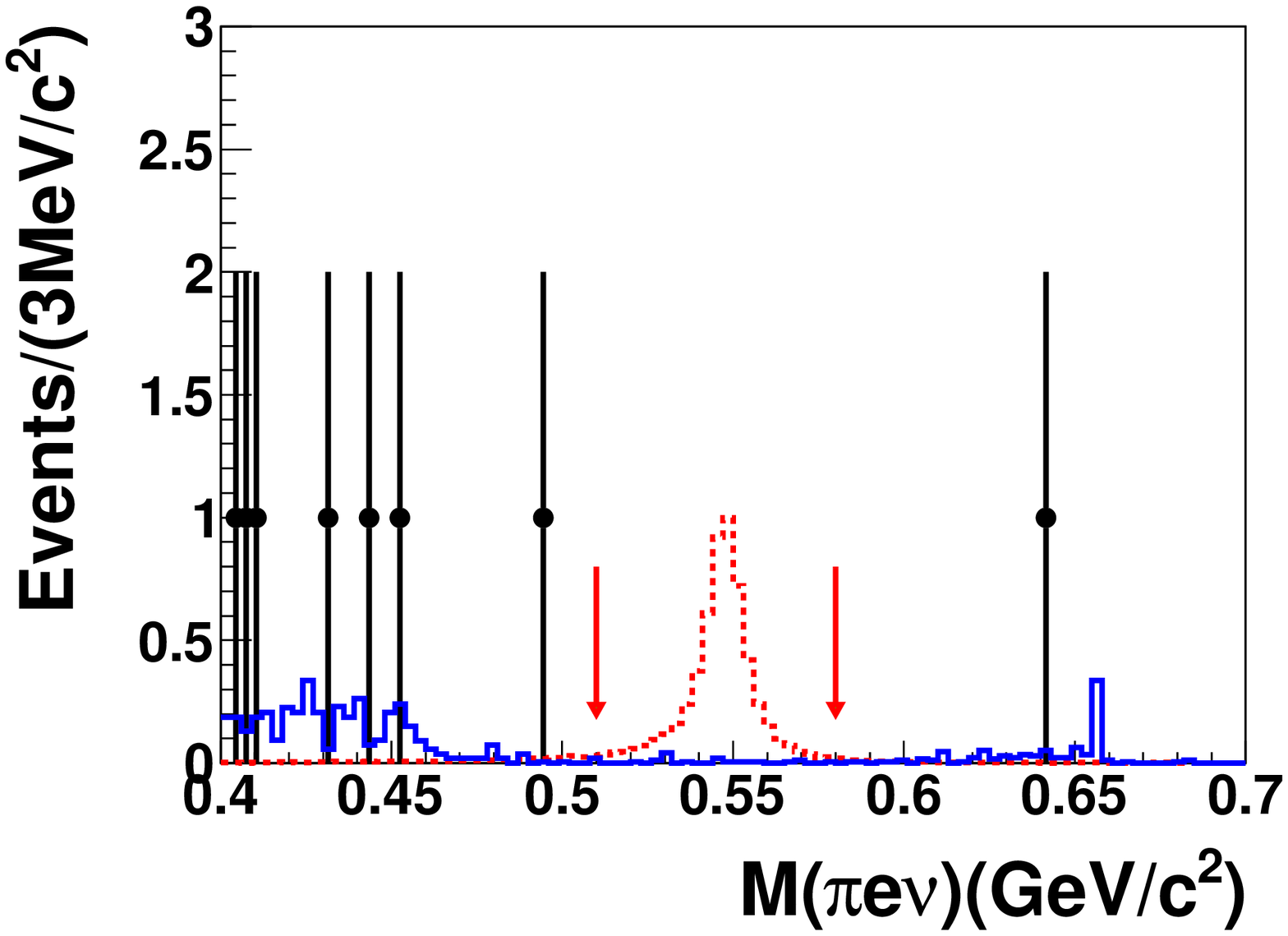}}
  \put(-90,72) {\bf \scriptsize (a)}
  {\label{fig:subfig:b}\includegraphics[width=0.48\columnwidth]{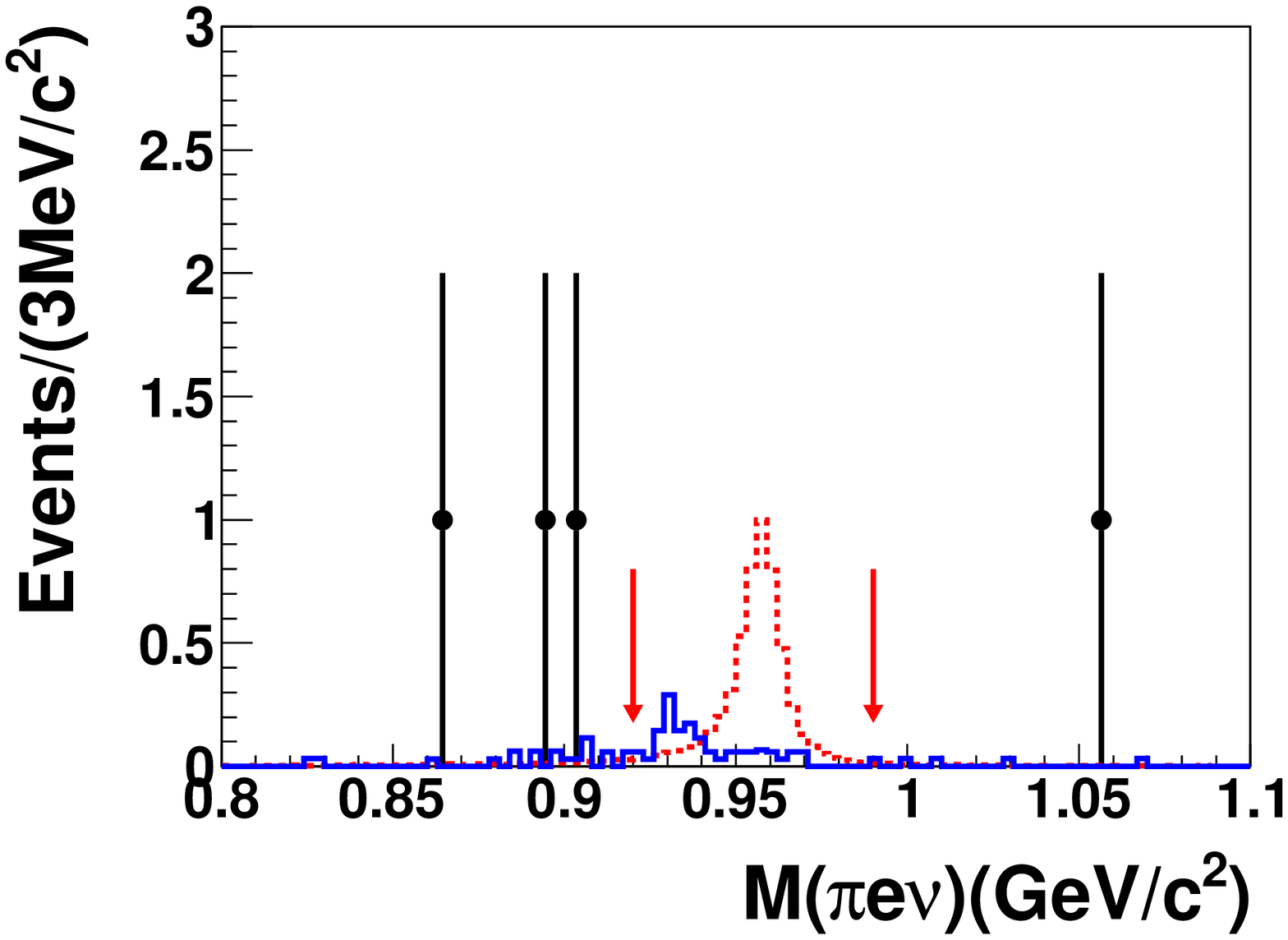}}
  \put(-90,72) {\bf \scriptsize (b)}
\caption{The $m_{\pi^+ e^- \bar{\nu}_e}$ distributions of candidate
events:
 (a) for $J/\psi \to \phi \eta$ ($\eta \to \pi^+ e^- \bar{\nu}_e$); (b) for $J/\psi \to \phi
 \eta^\prime$ ($\eta^\prime\to \pi^+ e^- \bar{\nu}_e$).  For both (a) and (b): the data (dots with error bars) are
compared to the signal MC samples (red dashed histogram) and the
expected backgrounds (solid blue histogram). The arrows on the plots
indicate the signal regions of $\eta$ and $\eta^\prime$ candidates.}
\label{eta}
\end{figure}

Figures~\ref{phie} (a) and (b) show the invariant mass distribution
of $K^+K^-$ candidates, $m_{K^+K^-}$, after the above selections.
Clear  $\phi$ signals are seen. The invariant mass of $\pi^+ e^-
\bar{\nu}_e$  can be obtained as $m_{\pi^+ e^- \bar{\nu}_e} =
\sqrt{(E_{\pi^+} + E_{e^-} + E_{\bar{\nu}})^2 - (\vec{P}_{\pi^+} +
\vec{P}_{e^-} + \vec{P}_{\bar{\nu}})^2}$, where $E_{\bar{\nu}}\equiv
E_{\text{miss}}=|\vec{P}_{\text{miss}}|$ and $\vec{P}_{\bar{\nu}} =
\vec{P}_{\text{miss}}$.  Figures~\ref{eta} (a) and (b) show the
$m_{\pi^+ e^- \bar{\nu}_e}$ distributions for events with $1.01 <
m_{K^{+}K^{-}} < 1.03$ GeV/$c^{2}$ for the decays $J/\psi \to \phi
\eta$ ($\eta \to \pi^+ e^- \bar{\nu}_e $) and $J/\psi \to \phi
 \eta^\prime$ ($\eta^\prime\to \pi^+ e^- \bar{\nu}_e$), respectively. No events are observed in the $\eta$ and
$\eta^\prime$ signal regions. The signal regions for $\eta$ and
$\eta^\prime$ are defined in the ranges [0.51, 0.58] and [0.92,
0.99] GeV/$c^2$, respectively, on the mass of $\pi^+ e^- \bar{\nu}_e
$.

We use MC simulated events to determine selection efficiencies for
the signal channels and study possible backgrounds. With phase space
MC simulations, we obtain efficiencies of 17.9\% and 17.4\% for
$\eta$ and $\eta^\prime$ semileptonic decays, respectively.
According to the study of the $J/\psi$ inclusive MC sample, more
than 20 exclusive decay modes are identified as potential background
modes, and are studied with full MC simulations in order to
understand the backgrounds. The sources of backgrounds are divided
into two classes. Class I: The background is from $J/\psi \to \phi
\eta (\eta^\prime)$, $\phi\to K^+K^-$ and $\eta$ ($\eta^\prime$)
decays into other modes than the signal final states. We find that
the expected number of background events from this class is $0.18\pm
0.05$ ($0.58 \pm 0.09$) in the signal region for  $\eta$
($\eta^\prime$).   Class II: It is mainly from $J/\psi$ decays to
the final states without $\eta$ or $\eta^\prime$, such as $\phi
\pi^+\pi^-$, $\phi f_0(980)$ ($f_0(980)\to \pi^+ \pi^-$), and
$K^{*0}\bar{K}^{*0}$ ($K^{*0} \to K^{\pm} \pi^\mp$). The expected
number of background events from class II is $0.05\pm 0.04$
($0.45\pm 0.13$) in the signal region for the $\eta$ ($\eta^\prime$)
case.  The total number of background events is $0.23\pm 0.06$
($1.03\pm 0.16$) in the signal region for $\eta$ ($\eta^\prime$).


After all selection criteria are applied, no event survives in the
$\eta$ and $\eta^\prime$ signal regions. The signal components and
the expected background shapes are projected and compared to data
for both $\eta$ and $\eta^\prime$ cases, as shown in Figs.~\ref{eta}
(a) and (b). We set an upper limit at the 90\% confidence level
(C.L.) to be $N^\eta_{\text{UL}}=2.36$ ($
N^{\eta^\prime}_{\text{UL}}=1.59$) for $\eta$($\eta^\prime$), using
the POLE$^{++}$ program~\cite{pole} with the Feldman-Cousins
frequentist approach~\cite{FC}. The information used to obtain the
upper limit includes the number of observed events in the signal
region, and the expected number of background events and their
uncertainty.


\subsection{Analyses for $\eta~(\eta^\prime) \to \pi^+ \pi^- \pi^0 (\eta) $  }
\label{sec:gg:selection}

 The $\eta \to \pi^+\pi^- \pi^0$ and $\eta^\prime \to \pi^+\pi^-
 \eta$ decays
 are also studied
in  $J/\psi \to \phi \eta$ and $\phi \eta^\prime$ processes, in
order to obtain the ratio of ${\mathcal B}(\eta(\eta^\prime)\to
\pi^+ e^-\bar{\nu}_e +c.c.)$ to ${\mathcal B}(\eta\to
\pi^+\pi^-\pi^0)$ (${\mathcal B}(\eta\to \pi^+\pi^-\eta))$. The
advantage of measuring the ratios of semileptonic weak decays over
hadronic decays $\frac{{\mathcal B}(\eta\to \pi^+ e^- \bar{\nu}_e
+c.c.)}{{\mathcal B}(\eta \to \pip\pim\piz)}$ and $\frac{{\mathcal
B}(\eta^\prime\to \pi^+ e^-\bar{\nu}_e +c.c.)}{{\mathcal
B}(\eta^\prime \to \pip\pim\eta)}$ is that the uncertainties due to
the total number of $J/\psi$ events, tracking efficiency, PID for
kaon and one pion, the number of the charged tracks, and residual
noise in the EMC cancel.

The selection criteria for the charged tracks are the same as those
for the $J/\psi \to \phi \eta~(\eta^\prime)$, $\eta~(\eta^\prime)\to
\pi^+ e^-\bar{\nu}_e $ decays except for the electron identification
requirement. The candidate events are required to have two charged
kaons and two charged pions with opposite charge. In addition, two
photon candidates are required to reconstruct $\pi^0 \to \gamma
\gamma$ and $\eta \to \gamma \gamma$ in the $\eta \to
\pi^+\pi^-\pi^0$ and $\eta^\prime \to \pi^+\pi^-\eta$ decays,
respectively. The photon candidates are required to be isolated from
all charged tracks by more than $10^\circ$ which is different from
the selection criteria for the $J/\psi \to \phi \eta~(\eta^\prime)$,
$\eta~(\eta^\prime)\to \pi^+ e^-\bar{\nu}_e $ decays in order to
improve the efficiency of $\pi^0(\eta)$ reconstruction.
 A four-constraint (4C) energy-momentum conservation
kinematic fit is performed to the $J/\psi \to K^+K^-\pi^+\pi^-
\gamma \gamma$ hypothesis, and only events with \chisq{4C} $<200$
 are accepted. For events with more than two photon
candidates, the combination with the minimum $\chi^2_{4C}$ is
selected. After the 4C fits, the $\pi^0$ and $\eta$ signal windows
on the $\gamma \gamma$ invariant mass distributions are defined in
the ranges $0.115 <m_{\gamma\gamma} < 0.150$ GeV/$c^2$ and $0.518
<m_{\gamma\gamma} < 0.578$ GeV/$c^2$, respectively.

\begin{figure*}[hbtp]
\begin{center}
  \includegraphics[width=0.9\columnwidth]{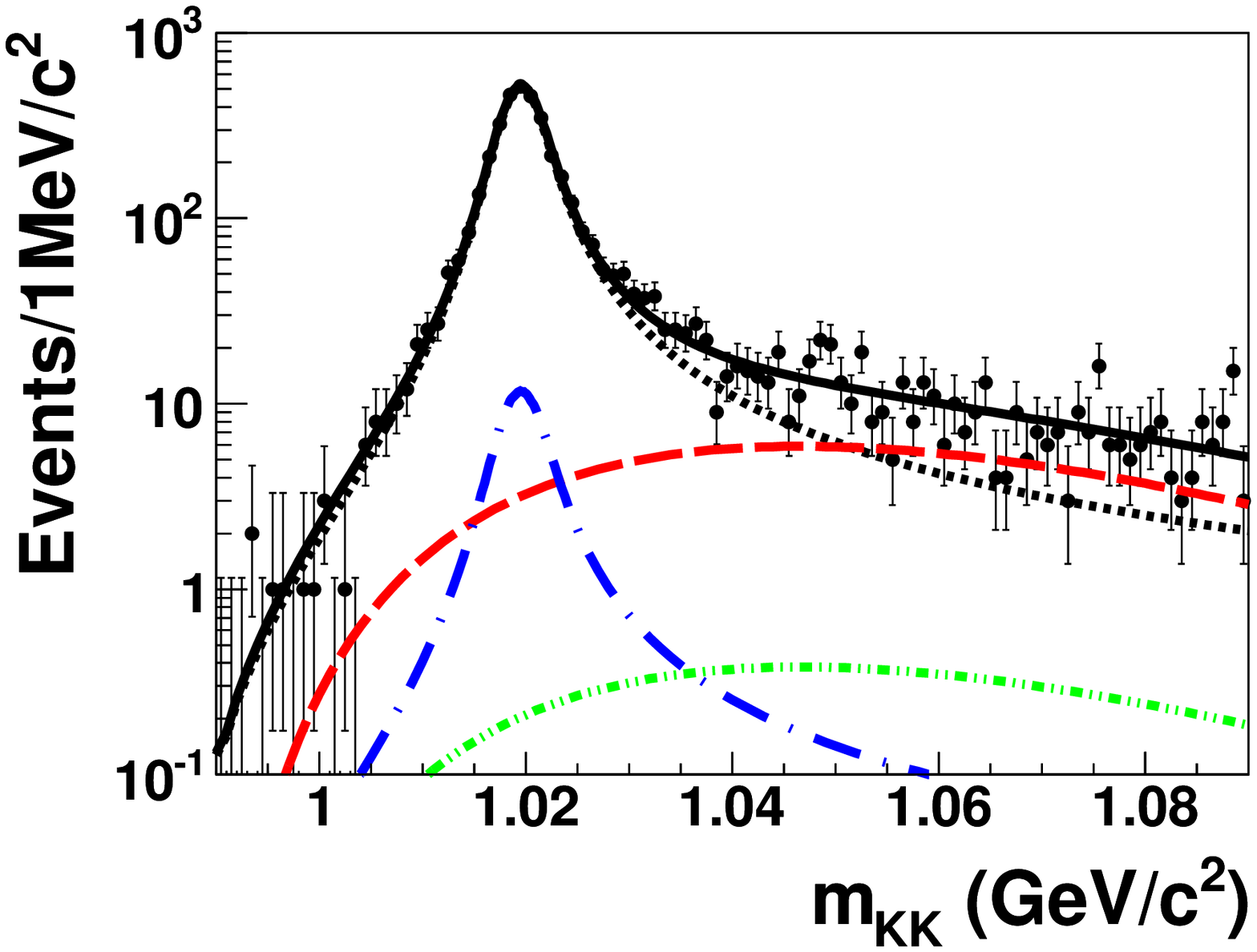}\put(-170,132) {\bf (a)}
  \includegraphics[width=0.9\columnwidth]{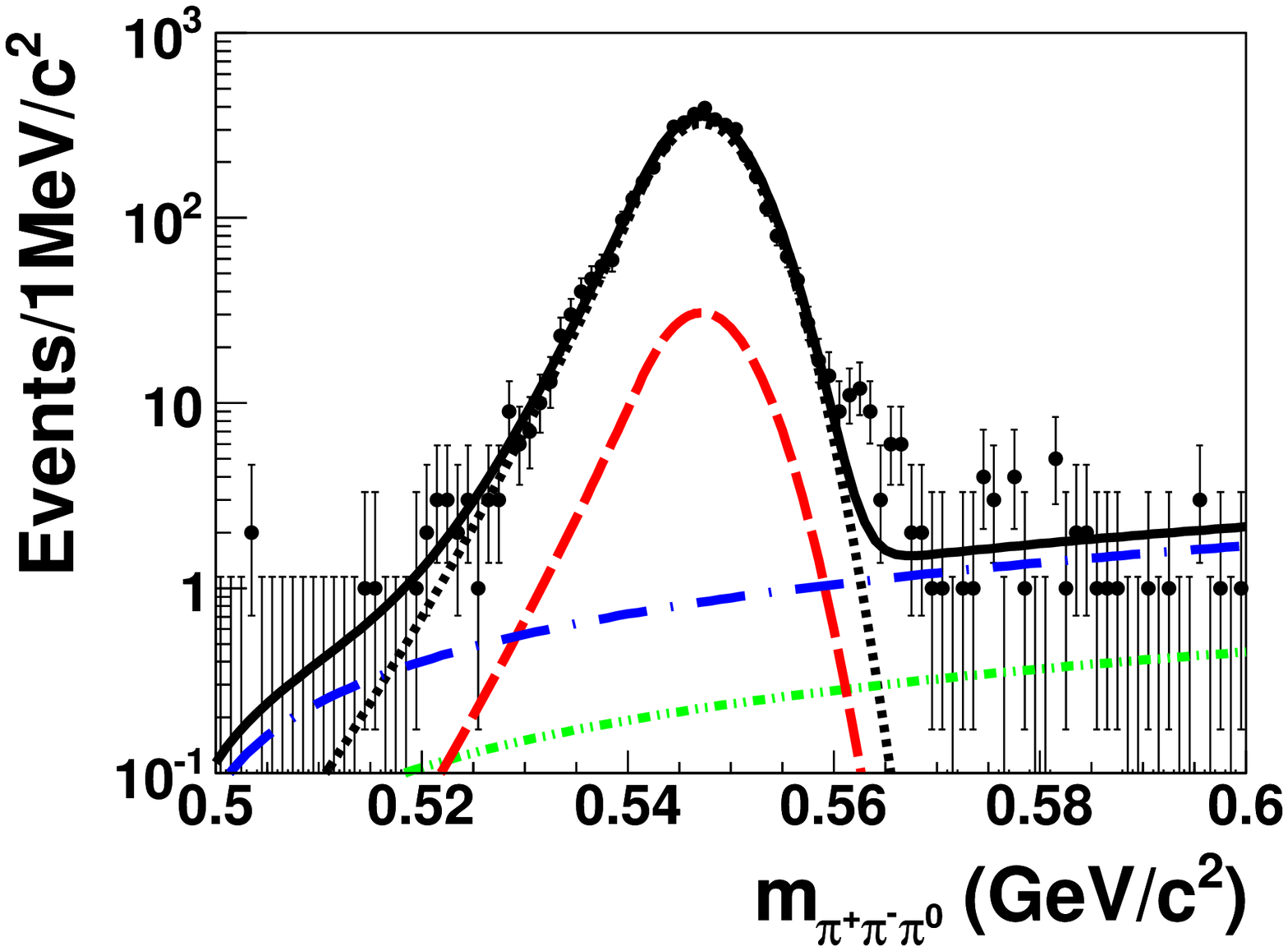}\put(-170,132) {\bf (b)}
  \caption{The (a) $m_{KK}$ and (b) $m_{\pi^+\pi^-\pi^0}$
  distributions with fit results superimposed for $J/\psi \to \phi \eta$, $\phi \to K^+K^-$, $\eta\to \pi^+\pi^-\pi^0$. Points with error bars are
  data. The (black) solid curves show the results of the total fits, and the (black)
  short-dashed curves are for signals. The (blue)
dotted-dash curve shows non-$\eta$-peaking backgrounds,  the (red)
dashed curve shows the non-$\phi$-peaking background, and the
(green) dotted-short-dash curve shows non-$\phi\eta$-peaking
backgrounds.}
  \label{figetafit}
\end{center}
\end{figure*}

\begin{figure*}[hbtp]
\begin{center}
  \includegraphics[width=0.9\columnwidth]{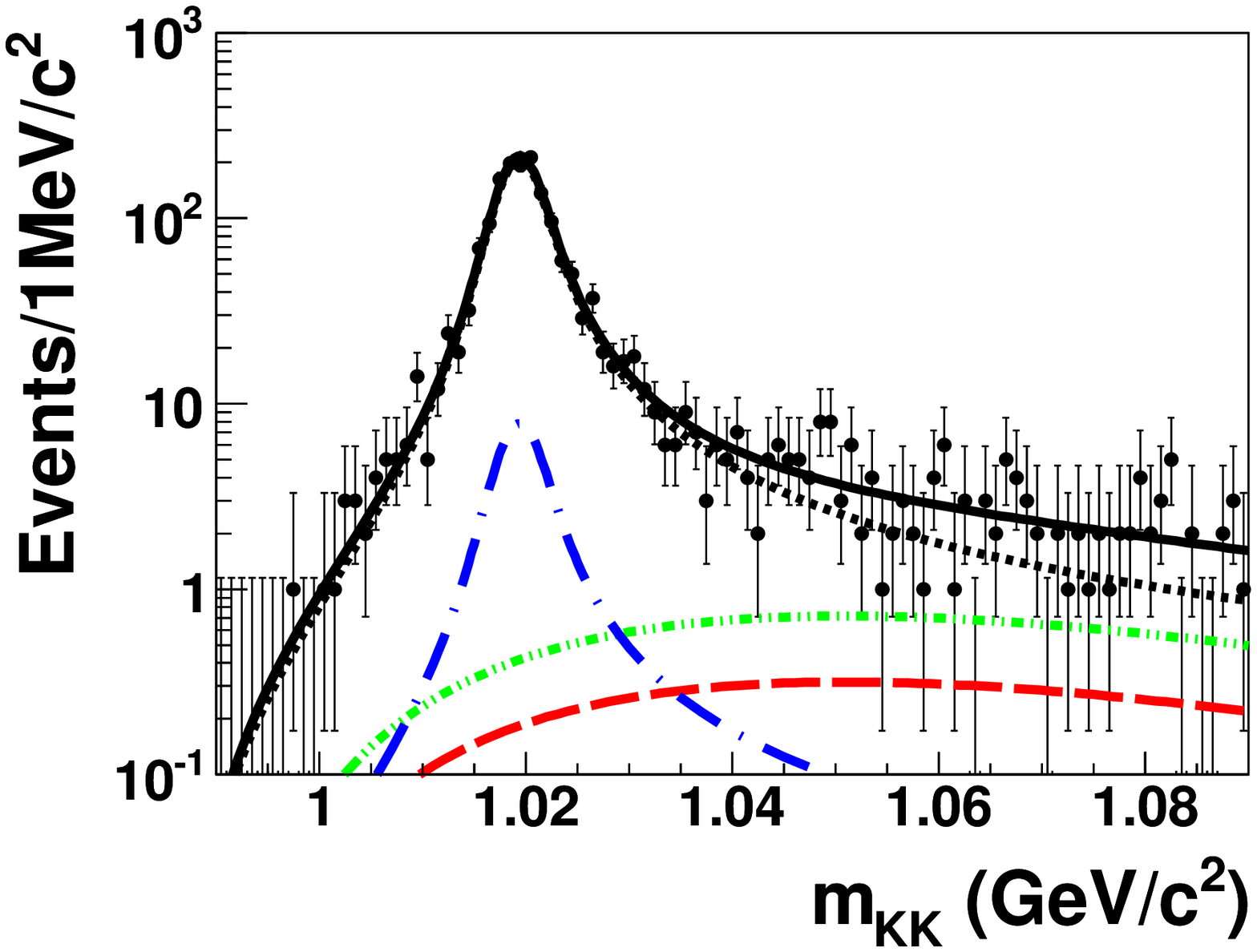}\put(-170,132) {\bf  (a)}
  \includegraphics[width=0.9\columnwidth]{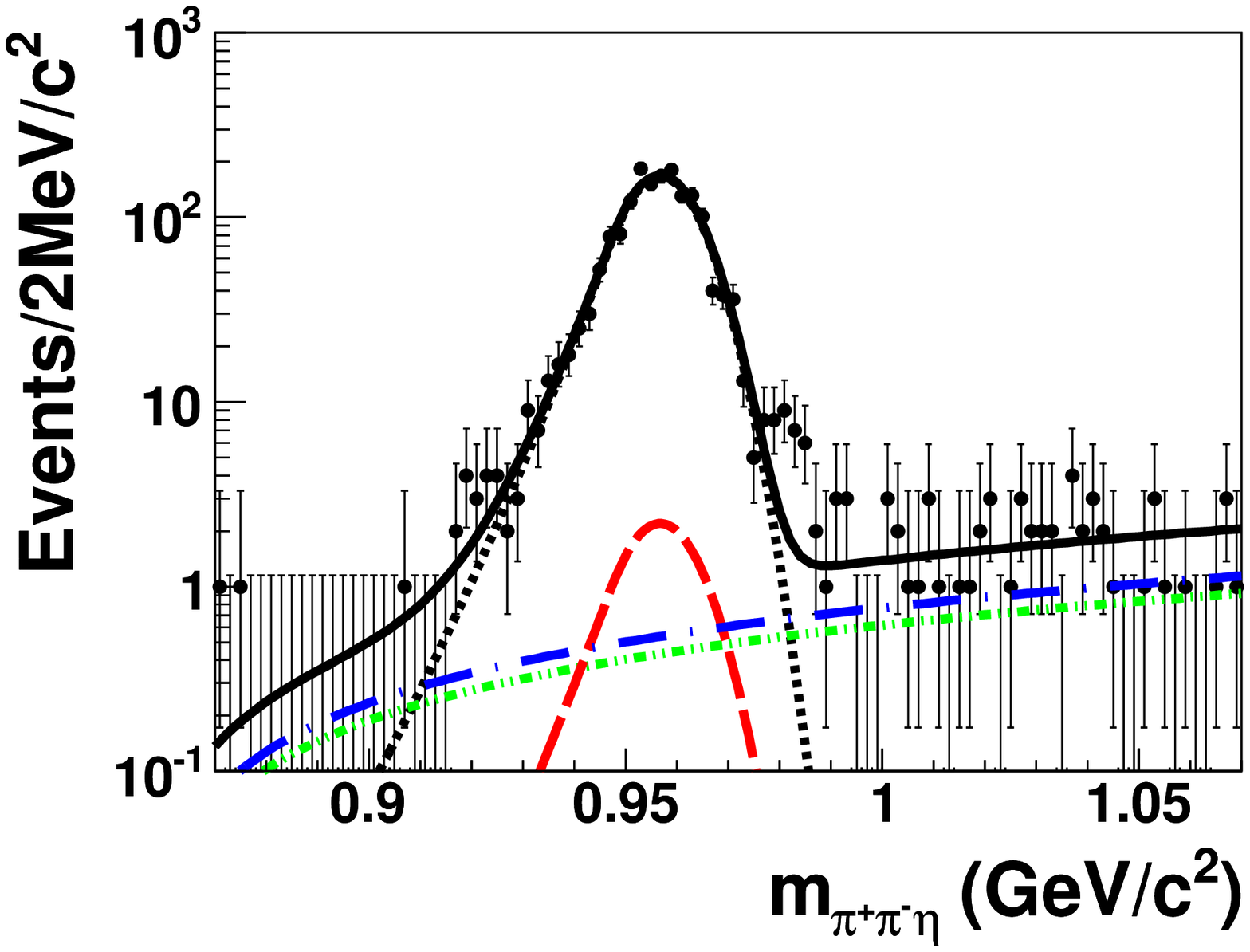}\put(-170,132) {\bf  (b)}
  \caption{The (a) $m_{KK}$ and (b) $m_{\pi^+\pi^-\eta}$
  distributions with fit results superimposed for
  $J/\psi \to \phi \eta^\prime$, $\phi \to K^+K^-$, $\eta^\prime\to \pi^+\pi^-\eta$.
  Points with error bars are
  data. The (black) solid curves show the results of the total fits, and the (black) short-dashed
   curves are for signals.  The (blue)
dotted-dash curve shows non-$\eta^\prime$-peaking backgrounds,  the
(red) dashed curve shows the non-$\phi$-peaking background, and the
(green) dotted-short-dash curve shows non-$\phi\eta^\prime$-peaking
backgrounds}
  \label{figetapfit}
\end{center}
\end{figure*}

The numbers of $J/\psi\to\phi\eta~(\eta^\prime),$
$\eta~(\eta^\prime)\to\pi^+\pi^-\pi^0(\eta)$ events are obtained
from an unbinned extended maximum likelihood (ML) fit to the
$K^{+}K^{-}$ {\it versus} $\pi^+\pi^-\pi^0(\eta)$ invariant mass
distributions. The projection of the fit on the $m_{KK}$
($m_{\pi^+\pi^-\pi^0}$ and $m_{\pi^+\pi^-\eta}$) axis is shown in
Figs.~\ref{figetafit} (a) and~\ref{figetapfit} (a)
[Figs.~\ref{figetafit} (b) and ~\ref{figetapfit} (b)] for the $\eta$
and $\eta^\prime$ cases, respectively. In the ML fits, we require
that $0.99$ GeV/$c^2$ $<m_{KK} <1.09$ GeV/$c^2$ and $0.50$ GeV/$c^2$
$< m_{\pi^+\pi^-\pi^0}<0.60$ GeV/$c^2$ ($0.87$ GeV/$c^2$ $<
m_{\pi^+\pi^-\eta}<1.07$ GeV/$c^2$) for the $\eta~(\eta^\prime)$
case. The signal shape for $\phi$ is modeled with a relativistic
Breit-Wigner ($RBW$) function~\cite{rbw} convoluted with a Gaussian
function that represents the detector resolution. In the fit, the
width of $\phi$ is fixed at the PDG value, and its central mass
value is floated, and the width of the Gaussian is free; the signal
shape for $\eta~(\eta^\prime)$ is described by a Crystal Ball ($CB$)
function~\cite{cbf}, and its parameters are floated. The backgrounds
are divided into three categories:
non-$\phi\eta(\eta^\prime)$-peaking background ({\it i.e.,} $J/\psi
\to \pi^+\pi^-\pi^0 K^+K^-$); non-$\phi$-peaking background [{\it
i.e.,} $J/\psi \to K^+K^- \eta~(\eta^\prime)$]; and
non-$\eta(\eta^\prime)$-peaking background ({\it
  i.e.,} $J/\psi \to \phi \pi^+\pi^-\pi^0$). The
probability density functions (PDF) for non-$\phi$-peaking background
in the $m_{KK}$ distribution is parameterized by~\cite{kkbgfunction}
\begin{eqnarray}
B(m_{KK}) = (m_{KK}-2m_{K})^{a} \cdot e^{-bm_{KK}-cm^{2}_{KK}},
\label{bgphipdf}
\end{eqnarray}
where $a$, $b$ and $c$ are free parameters, and $m_K$ is the nominal
mass value of the charged kaon from the PDG~\cite{pdg}. The shape
for the non-$\eta(\eta^\prime)$-peaking background in the
$m_{\pi^+\pi^-\pi^0 (\eta)}$ distribution is modeled by a
first-order Chebychev polynomial function [$B(m_{\pi^+\pi^-\pi^0
(\eta)})$]. All parameters related to the background shape are
floated in the fit to data.  Totally, 14 parameters including signal
and background yields are floated in the fit.  The PDFs for signal
and backgrounds are combined in the likelihood function
$\mathcal{L}$, defined as a function of the free parameters
$N^\eta$, $N^{\text{non-}\phi\eta}_{\text{bkg}}$,
$N^{\text{non-}\phi}_{\text{bkg}}$, and
$N^{\text{non-}\eta}_{\text{bkg}}$:
\begin{eqnarray}
\mathcal{L} &=& \frac{e^{-(N^\eta
+N^{\text{non-}\phi\eta}_{\text{bkg}}+N^{\text{non-}\phi}_{\text{bkg}}+N^{\text{non-}\eta}_{\text{bkg}})}}
{N!} \nonumber \\
&&\times \prod^N_{i=1}[N^\eta RBW(m^i_{KK})\times
CB(m^i_{\pi^+\pi^-\pi^0}) \nonumber\\
&&+ N^{\text{non-}\phi\eta}_{\text{bkg}} B(m^i_{KK})\times
B(m^i_{\pi^+\pi^-\pi^0}) \nonumber\\
&&+ N^{\text{non-}\phi}_{\text{bkg}} B(m^i_{KK})\times CB(m^i_{\pi^+\pi^-\pi^0}) \nonumber\\
&&+ N^{\text{non-}\eta}_{\text{bkg}} RBW(m^i_{KK})\times
B(m^i_{\pi^+\pi^-\pi^0})], \label{eq:pdf:2d:gg}
\end{eqnarray}
where $N^\eta$ is the number of $J/\psi \to \phi \eta,~\phi \to
K^+K^-,~\eta \to \pi^+\pi^-\pi^0$ events, and
$N^{\text{non-}\phi\eta}_{\text{bkg}}$,
$N^{\text{non-}\phi}_{\text{bkg}}$, and
$N^{\text{non-}\eta}_{\text{bkg}}$ are the numbers of the
corresponding three kinds of backgrounds. The fixed parameter $N$ is
the total number of selected events in the fit region, and
$m^i_{KK}$ ($m^i_{\pi^+\pi^-\pi^0}$) is the value of $m_{KK}$
($m_{\pi^+\pi^-\pi^0}$) for the $i$th event. We use the product of
the PDFs, since we have verified that $m_{KK}$ and
$m_{\pi^+\pi^-\pi^0}$ are uncorrelated for each component. The
negative log-likelihood ($-\text{ln}\mathcal{L}$) is then minimized
with respect to the extracted yields.   The resulting fitted signal
and background yields are summarized in Table~\ref{tab:ggresults}.
We also obtain the results for the $\eta^\prime$ case by replacing
$\eta$ and $\pi^0$ with $\eta^\prime$ and $\eta$ in
Eq.~(\ref{eq:pdf:2d:gg}).
  The fitted results for $\eta~(\eta^\prime)
\to \pi^+\pi^-\pi^0(\eta)$ are shown in Fig.~\ref{figetafit}
(Fig.~\ref{figetapfit}). The detection efficiencies are determined
with MC simulations to be $20.37\%$ and $20.89\%$ for $\eta$ and
$\eta^\prime$, respectively.
\begin {table}[htbp]
\begin {center}\small
\caption{The fitted signal and background yields for
$J/\psi\to\phi\eta~(\eta^\prime)$,
$\eta~(\eta^\prime)\to\pi^+\pi^-\pi^0 (\eta)$, and
$\epsilon^\eta(\epsilon^{\eta^\prime})$ is its selection efficiency.
} \label{tab:ggresults}
\renewcommand{\arraystretch}{1.4}
\begin {tabular}{l|cc}
\hline \hline
 & \multicolumn{2}{c}{Value} \\
 Quantity            & $\eta$                 & $\eta^\prime$  \\
\hline
$N^\eta (N^{\eta^\prime})$            &    $3850 \pm 73$       &  $1623 \pm 44$    \\
\hline
$N^{\text{non-}\phi\eta}_{\text{bkg}}(N^{\text{non-}\phi\eta^\prime}_{\text{bkg}})$        &        $24 \pm 8$          &  $49 \pm 10$ \\
\hline
$N^{\text{non-}\phi}_{\text{bkg}}(N^{\text{non-}\phi}_{\text{bkg}})$                          &     $367 \pm 43$                  &  $22 \pm 17$  \\
\hline
$N^{\text{non-}\eta}_{\text{bkg}}(N^{\text{non-}\eta^\prime}_{\text{bkg}})$                &   $88 \pm 14$             &  $61 \pm 12$  \\
\hline
$\epsilon^\eta (\epsilon^{\eta^\prime})$                    &        20.37\%             &   20.89\%       \\
\hline\hline
\end {tabular}
\end {center}
\end {table}


\section{Systematic uncertainties }
\label{sec:sys:selection}

Contributions to the systematic error on the ratios are summarized
in Table~\ref{tab:syserr}. The uncertainty, due to the requirement
of no neutral showers in the EMC inside a cone spanning 0.3 (1.5)
rad around the direction of the missing momentum for
$\eta~(\eta^\prime)$ decay is obtained using the control sample of
decays $J/\psi\to \phi \eta^\prime$, $\phi\to K^+K^-$,
$\eta^{\prime}\to \gamma\pi^+\pi^-$. We calculated the missing
momentum of the $K^+K^-\pi^+\pi^-$ system, and define the same cones
around the direction of the missing momentum as in the $\eta$
($\eta^\prime$) semileptonic analysis. The ratios of events with the
requirement on the number of extra photons to events without the
requirement are obtained for both data and MC simulation. The
difference 0.1\% (1.1\%) is considered as a systematic error for the
$\eta$ ($\eta^\prime$) case.
%

We also use the control sample of $J/\psi\to \phi \eta^\prime$,
$\phi\to K^+K^-$, $\eta^{\prime}\to \gamma\pi^+\pi^-$ to obtain the
uncertainty due to the requirement on the missing momentum
$P_{\text{miss}}>0.03$ GeV/$c$ for both $\eta$ and $\eta^\prime$
cases. Thus we calculated the missing momentum of the
$K^+K^-\pi^+\pi^-$ system. The ratios of events with the requirement
on the missing momentum $P_{\text{miss}}>0.03$ GeV/$c$ to events
without the requirement are obtained for both data and MC
simulation. The difference 2.5\% is considered as a systematic error
for both $\eta$ and $\eta^\prime$ cases.

The phase space MC is used to generate  $\eta~(\eta^\prime) \to
\pi^+e^-\bar{\nu}_e$ decays. In Ref.~\cite{neufeld}, the transition
form factors $f^{\eta\pi}_{\pm}$ are calculated at the one-loop
level in the chiral perturbation theory. We use the model
predictions to generate signal MC events, and find that the
uncertainty on the detection efficiency is changed by 1.0\% (5.0\%)
for the $\eta$ ($\eta^\prime$) case.

Since the uncertainties on the PID of the electron and one of the
pions do not cancel in the ratio, the efficiencies for pion and
electron PID are obtained with the control samples of  $J/\psi \to
\pi^+\pi^-\pi^0$ and radiative Bhabha scattering $e^+e^- \to \gamma
e^+e^-$ (including $J/\psi\to \gamma e^+e^-$), respectively. Samples
with backgrounds less than 1.0\% are obtained~\cite{xugf}. The
differences between data and MC for the efficiencies of pion and
electron PID  are about 1.0\% and 1.2\%, respectively, which are
taken as systematic errors. Using the same control samples, we
estimate the uncertainty due to the requirement of $E/p$ for the
electron selection to be 3.5\% (3.4\%) for the $\eta$
($\eta^\prime$) case, and the uncertainty due to the requirement of
$E/p$ for pion selection in the $\eta$ semileptonic decay is
estimated to be 0.8\%. The systematic uncertainty due to the
requirements of $\phi$ and $\eta$ ($\eta^\prime$) mass windows are
estimated to be 1.4\% and 0.04\% (0.2\%) by using the control sample
of $J/\psi\to\phi\eta~(\eta^\prime)$,
$\eta~(\eta^\prime)\to\pi^+\pi^-\pi^0(\eta)$.

The uncertainty in the determination of the numbers of observed
events for $J/\psi \to \phi \eta$ [$\eta \to
\pi^+\pi^-\pi^0(\pi^0\to \gamma\gamma)$] and $J/\psi \to \phi
\eta^\prime$ [$\eta^\prime \to \pi^+\pi^-\eta~(\eta\to
\gamma\gamma)$] decays are estimated on the basis of earlier
published results. The photon detection efficiency and its
uncertainty are studied by three different methods in
Ref~\cite{xugf}. The systematic error of photon detection is
estimated to be $1.0\%$ per photon. In the fit to the $\phi$ mass
distribution, the mass resolution is fixed to the MC simulation; the
level of possible discrepancy is determined with a smearing
Gaussian, for which a non-zero $\sigma$ would represent a MC-data
difference in the mass resolution. The uncertainty associated with
the difference determined in this way is 0.03\% (0.06\%) for the
$\eta~(\eta^\prime)$ case. The systematic uncertainty due to the
choice of parameterization for the shape of the
non-$\phi\eta(\eta^\prime)$-peaking background is estimated by
varying the order of the polynomial in the fit; we find the relative
changes on the $\eta~(\eta^\prime)$ signal yield of 1.3\%~(0.8\%),
which is taken as the uncertainty due to the background shapes. The
systematic errors from $\pi^0$ ($\eta$) reconstruction from $\gamma
\gamma$ decays is determined to be 1.0\% per $\pi^0$ ($\eta$) by
using a high purity control sample of $J/\psi \to \pi^0 p\bar{p}$
($J/\psi \to \eta p\bar{p}$) decay~\cite{lei}. The branching
fractions for the $\pi^0$ and $\eta\to\gamma\gamma$ decays are taken
from the PDG~\cite{pdg}. The uncertainties on the branching
fractions are taken as a systematic uncertainty in our measurements.
The total systematic error $\sigma^{sys}_{\eta}$
($\sigma^{sys}_{\eta^\prime}$) on the ratio is 5.6\% (7.4\%) for
$\eta$ ($\eta^\prime$), as summarized in Table~\ref{tab:syserr}.
\begin{table}[htb]
  \caption{Summary of relative systematic errors for the determination
  of ratios of semileptonic over hadronic decays. The first
    nine lines are relevant for the semileptonic weak decay chain
    $J/\psi \rightarrow \phi\eta~(\eta^\prime)$,
    $\eta~(\eta^\prime)\to \pi^+ e^-\bar{\nu}_e$. The next five
    lines are relevant for the determination of the signal yield of the hadronic
    decay process $J/\psi \rightarrow \phi \eta~(\eta^\prime)$,  $\eta\to\pi^+\pi^-\pi^0$
    ($\eta^\prime\to\pi^+\pi^-\eta$).}
  \begin{tabular}{l|cc}
    \hline \hline
    & \multicolumn{2}{|c}{Sys. error(\%)} \\
    Sources& $\eta$ & $\etap$ \\ \hline
    Requirement on $N_{\text{shower}}$  & 0.1 & 1.1 \\
    Requirement on $P_{\text{miss}}$ & 2.5 & 2.5 \\
    Signal model  & 1.0 & 5.0 \\
    Electron PID & 1.2  & 1.2 \\
    Requirement on $E/p$ for $e$ & 3.5 & 3.4 \\
     Requirement on $E/p$ for $\pi$   & 0.8 & - \\
    $\phi$ mass window              &  1.4          &     1.4           \\
    $\eta$ ($\eta^\prime$) mass window & 0.0 & 0.2\\
    \hline
    Photon efficiency & 2.0 & 2.0 \\
    $\pi$ PID                & 1.0 & 1.0 \\
    Signal shapes for  $\eta(\eta^\prime)\to \pi^+\pi^-\pi^0(\eta)$            &  0.0        &  0.1    \\
Background shape for $\eta(\eta^\prime)\to \pi^+\pi^-\pi^0(\eta)$  &   1.3    &  0.8   \\
    $\pi^0(\eta)$ reconstruction & 1.0 & 1.0 \\
     Cited ${\mathcal B}(\pi^0(\eta)\to \gamma\gamma)$ & 0.0 & 0.5  \\ \hline
     Total & 5.6 & 7.4 \\ \hline \hline
  \end{tabular}
  \label{tab:syserr}
\end{table}

\section{Results}

The upper limit on the ratio of branching fractions of the semileptonic decay $\mathcal{B}(\eta\to
\pi^+e^-\bar{\nu}_e +c.c.)$ over the hadronic decay $\mathcal{B}(\eta\to\pi^+\pi^-\pi^0)$ is
calculated with
\begin{eqnarray}
  \frac{\mathcal{B}(\eta\to\pi^+e^-\bar{\nu}_e +c.c.)}{\mathcal{B}(\eta\to\pi^+\pi^-\pi^0) } <
  \frac{N^{\eta}_{\text{UL}}/\epsilon^{\text{SL}}_\eta}{N^{\eta}/\epsilon^\eta}\,
  \frac{\mathcal{B}(\pi^0\to \gamma\gamma)}{(1-\sigma_\eta)}\,,
  \label{eq:upper}
\end{eqnarray}
where $N^{\eta}_{\text{UL}}$ is the 90\% upper limit of the observed
number of events for $J/\psi \to \phi \eta$, $\phi\to K^+K^-$,
$\eta\to \pi^+ e^-\bar{\nu}_e $ decay, $\epsilon^{\text{SL}}_{\eta}$
is the MC determined efficiency for the signal channel, $N^\eta$ is
the number of events for the $J/\psi \to \phi \eta$, $\phi\to
K^+K^-$, $\eta \to \pi^+\pi^-\pi^0(\pi^0\to \gamma\gamma)$ decay,
$\epsilon^\eta$ is the MC determined efficiency for the decay mode,
and $\sigma_{\eta}=
\sqrt{(\sigma^{\text{sys}}_{\eta})^2+(\sigma^{\text{stat}}_{\eta})^2}
= 5.9\%$, where $\sigma^{\text{sys}}_{\eta}$ and
$\sigma^{\text{stat}}_{\eta}$ are the total relative systematic
error for the $\eta$ case from Table~\ref{tab:syserr} and the
relative statistical error of $N^\eta$, respectively. For
$\eta^\prime$,  $\sigma_{\eta^\prime}=
\sqrt{(\sigma^{\text{sys}}_{\eta^\prime})^2+(\sigma^{\text{stat}}_{\eta^\prime})^2}=7.9\%$.
The relative statistical error of $N^\eta$ ($N^{\eta^\prime}$) is
1.9\% (2.7\%). We also obtain the upper limit on the ratio of
$\mathcal{B}(\eta^\prime \to\pi^+e^-\bar{\nu}_e +c.c.)$ to
$\mathcal{B}(\eta^\prime \to\pi^+\pi^-\eta)$ by replacing $\eta$ and
$\mathcal{B}(\pi^0\to\gamma\gamma)$ with $\eta^\prime$ and
$\mathcal{B}(\eta\to\gamma\gamma)$, respectively,  in
Eq.~(\ref{eq:upper}). Since only the statistical error is considered
when we obtain the 90\% upper limit of the number of events, to be
conservative, $N^\eta_{\text{UL}}$ and $N^{\eta^\prime}_{\text{UL}}$
are shifted up by one sigma of the additional uncertainties
($\sigma_\eta$ or $\sigma_{\eta^\prime}$).

\begin{table}[htb]
  \caption{The numbers used in the calculations of the ratios in
    Eq.~(\ref{eq:upper}), where $N^\eta_{\text{UL}}(N^{\eta^\prime}_{\text{UL}})$ is
    the upper limit of the signal events at the 90\% C.L.,
    $\epsilon^{\text{SL}}_\eta(\epsilon^{\text{SL}}_{\eta^\prime})$
    is the selection efficiency,
    $N^\eta(N^{\eta^\prime})$ is the
    number of the events of $J/\psi\to\phi\eta(\eta^\prime)$, $\phi\to
    K^+ K^-$, $\eta\to\pi^+\pi^-\pi^0$ and $\pi^0\to\gamma\gamma$
    ($\eta^\prime\to\pi^+\pi^-\eta$ and $\eta\to\gamma\gamma$)
    , $\epsilon^\eta$ ($\epsilon^{\eta^\prime}$) is its selection efficiency, $\sigma_\eta^{\text{stat}}$
    ($\sigma_{\eta^\prime}^{\text{stat}}$) is the relative statistical error of
    $N^\eta(N^{\eta^\prime})$, and
    $\sigma_\eta(\sigma_{\eta^\prime})$ is the total relative error.}
  \begin{tabular}{l|cc}
    \hline \hline
    Quantity & \multicolumn{2}{|c}{Value} \\
    & $\eta$ & $\etap$ \\\hline
    $N^{\eta}_{UL}$ ($N^{\eta^\prime}_{UL}$) & 2.36 & 1.59 \\
    $\epsilon^{\text{SL}}_\eta$ ($\epsilon^{\text{SL}}_{\eta^\prime}$) & 17.9\% & 17.4\% \\
    $N^{\eta}$ ($N^{\eta^\prime}$) & $3850 \pm 73$ & $1623 \pm 44$ \\
   $\epsilon^\eta$ ($\epsilon^{\eta^\prime}$) & 20.37\% & 20.89\% \\
    $\sigma_\eta^{\text{stat}}$ ($\sigma_{\eta^\prime}^{\text{stat}}$) & 1.9\% & 2.7\% \\
    $\sigma_\eta$ ($\sigma_{\eta^\prime}$) & 5.9\% & 7.9\% \\
    \hline\hline
  \end{tabular}
  \label{tab:br}
\end{table}

Using the numbers in Table~\ref{tab:br}, the  upper limits on the
ratios $\frac{{\mathcal B}(\eta\to \pi^+ e^- \bar{\nu}_e
+c.c.)}{{\mathcal B}(\eta \to \pip\pim\piz)}$ and $\frac{{\mathcal
B}(\eta^\prime\to \pi^+ e^-\bar{\nu}_e +c.c.)}{{\mathcal
B}(\eta^\prime \to \pip\pim\eta)}$ are obtained at the 90\% C.L. of
$7.3\times 10^{-4}$ and $5.0\times 10^{-4}$, respectively.

\section{Summary}

A search for the semileptonic weak deacy modes
$\eta~(\eta^\prime)\to \pi^+e^-\bar{\nu}_e $ has been performed for
the first time in the process of $J/\psi \to \phi \eta
(\eta^\prime)$ using the $(225.3\pm 2.8)\times 10^6$  $J/\psi$
events measured at BESIII. We find no signal yields for the
semileptonic weak decays of $\eta$ and $\eta^\prime$. The upper
limits at the 90\% C.L. are $7.3\times 10^{-4}$ and $5.0\times
10^{-4}$ for the ratios of semileptonic over hadronic decay modes
$\frac{{\mathcal B}(\eta\to \pi^+ e^- \bar{\nu}_e +c.c.)}{{\mathcal
B}(\eta \to \pip\pim\piz)}$ and $\frac{{\mathcal B}(\eta^\prime\to
\pi^+ e^-\bar{\nu}_e +c.c.)}{{\mathcal B}(\eta^\prime \to
\pip\pim\eta)}$, respectively. The advantage of measuring the ratios
instead of the branching fractions of the semileptonic weak decays
is that many uncertainties cancel. Using the hadronic branching
fraction values of $\eta \to \pi^+\pi^-\pi^0$ and $\eta^\prime\to
\pi^+\pi^-\eta$ as listed by PDG~\cite{pdg}, we determine the
semileptonic decay rates to be ${\mathcal B}(\eta\to \pi^+ e^-
\bar{\nu}_e +c.c.)<1.7\times 10^{-4}$ and ${\mathcal
B}(\eta^\prime\to \pi^+ e^- \bar{\nu}_e +c.c.)<2.2\times 10^{-4}$ at
the 90\% C.L..

\begin{acknowledgements}
The BESIII collaboration thanks the staff of BEPCII and the
computing center for their hard efforts. This work is supported in
part by the Ministry of Science and Technology of China under
Contract No. 2009CB825200; National Natural Science Foundation of
China (NSFC) under Contracts Nos. 10625524, 10821063, 10825524,
10835001, 10935007, 11125525, 11061140514; Joint Funds of the
National Natural Science Foundation of China under Contracts Nos.
11079008, 11179007, 11179014; the Chinese Academy of Sciences (CAS)
Large-Scale Scientific Facility Program; CAS under Contracts Nos.
KJCX2-YW-N29, KJCX2-YW-N45; 100 Talents Program of CAS; Istituto
Nazionale di Fisica Nucleare, Italy; Ministry of Development of
Turkey under Contract No. DPT2006K-120470; U. S. Department of
Energy under Contracts Nos. DE-FG02-04ER41291, DE-FG02-94ER40823;
U.S. National Science Foundation; University of Groningen (RuG) and
the Helmholtzzentrum fuer Schwerionenforschung GmbH (GSI),
Darmstadt; WCU Program of National Research Foundation of Korea
under Contract No. R32-2008-000-10155-0; German Research Foundation
DFG within the Collaborative Research Center CRC1044.

\end{acknowledgements}

\end{document}